\documentclass[twocolumn,showpacs,preprintnumbers,amsmath,amssymb]{revtex4}

\usepackage{graphicx}
\usepackage{color}
\usepackage{dcolumn}
\usepackage{bm}

\begin{document}

\preprint{APS/123-QED}

\title{Structure, compressibility factor and dynamics of highly size-asymmetric binary hard-disk liquids}

\author{Wen-Sheng Xu, Zhao-Yan Sun\footnote{Correspondence author. E-mail: zysun@ciac.jl.cn}, and Li-Jia An}
\affiliation{State Key Laboratory of Polymer Physics and Chemistry,
Changchun Institute of Applied Chemistry, Chinese Academy of
Sciences, Changchun 130022, People's Republic of China}



\date{\today}

\begin{abstract}
By using event-driven molecular dynamics simulation, we investigate effects of varying the area fraction of the smaller component on structure, compressibility factor and dynamics of the highly size-asymmetric binary hard-disk liquids. We find that the static pair correlations of the large disks are only weakly perturbed by adding small disks. The higher-order static correlations of the large disks, by contrast, can be strongly affected. Accordingly, the static correlation length deduced from the bond-orientation correlation functions first decreases significantly and then tends to reach a plateau as the area fraction of the small disks increases. The compressibility factor of the system first decreases and then increases upon increasing the area fraction of the small disks and separating different contributions to it allows to rationalize this non-monotonic phenomenon. Furthermore, adding small disks can influence dynamics of the system in quantitative and qualitative ways. For the large disks, the structural relaxation time increases monotonically with increasing the area fraction of the small disks at low and moderate area fractions of the large disks. In particular, ``reentrant'' behavior appears at sufficiently high area fractions of the large disks, strongly resembling the reentrant glass transition in short-ranged attractive colloids and the inverted glass transition in binary hard spheres with large size disparity. By tuning the area fraction of the small disks, relaxation process for the small disks shows concave-to-convex crossover and logarithmic decay behavior, as found in other binary mixtures with large size disparity. Moreover, diffusion of both species is suppressed by adding small disks. Long-time diffusion for the small disks shows power-law-like behavior at sufficiently high area fractions of the small disks, which implies precursors of a glass transition for the large disks and a localization transition for the small disks. Therefore, our results demonstrate the generic dynamic features in highly size-asymmetric binary mixtures.
\end{abstract}


\maketitle

\section{Introduction}

The physical mechanism of the liquid-glass transition remains unclear despite its fundamental and technological importance~\cite{PWAnderson, Angell, Debenedetti, Berthier1}. Binary mixtures of particles, as a widely used model glass-former, play an important role in understanding the microscopic processes driving the glass formation during the past years. For example, the Kob-Andersen binary mixture has been used extensively in simulations to test theoretical predictions and analyze new phenomena within the glass community~\cite{Kob1, Kob2, Kob3, Berthier2, Berthier3}. To study the glass transition, binary mixtures are often introduced merely as a means of suppressing crystallization since glass cannot be formed in one-component systems with conventional isotropic pair potentials. However, recent studies show that mixing two constituents with different size not only offers a way to avoid crystallization but also possesses its own importance and exhibits very rich dynamical behavior~\cite{vanMegen, Sciortino1, Sciortino2, Voigtmann1, Voigtmann2, Voigtmann3, Voigtmann4, Moreno1, Moreno2, Weysser}.

For moderate disparate mixtures, in which structure and dynamics of the two species are qualitatively the same, generic mixing effects close to the glass transition have been identified~\cite{Sciortino1, Sciortino2, Voigtmann1, Voigtmann2}. Specifically, two qualitatively different scenarios for the structural relaxation have been demonstrated: Increasing the mixing percentage of small particles leads to a speed up of long-time dynamics for relatively large size disparity (say a size ratio $\delta=\sigma_{s}/\sigma_{l}=0.6$, where $\sigma_{s}$ and $\sigma_{l}$ indicate the diameters of small and large particles, respectively), while small disparity (e.g., $\delta=0.83$) leads to a slowing down\cite{Sciortino1}. Of special interest is the dynamic arrest in highly size-asymmetric binary mixtures, where the species' long-time dynamics can be quantitatively and qualitatively different. Moreno and Colmenero have demonstrated by molecular dynamics simulations~\cite{Moreno1, Moreno2} that a binary soft-sphere mixture with $\delta=0.4$ shows anomalous dynamic features, including the sublinear behavior for mean-squared displacements, concave-to-convex crossover for the intermediate scattering functions by varying temperature or wave vector and logarithmic decay for specific wave vectors of density-density correlators. These striking dynamic features differ from the standard pictures of structural dynamic arrest in glass-forming liquids and resemble the mode-coupling theory (MCT) predictions for fluids confined in matrices with interconnected voids~\cite{Krakoviack}. They also suggest two mechanisms of the dynamic arrest for different types of particles, i.e., dynamic arrest for the large particles originates from the competition between soft-sphere repulsion and depletion effects induced by neighboring small particles while for the small particles, it comes from the competition between bulk-like dynamics (induced by neighboring small particles) and confinement (induced by the slow matrix of large particles). Influence of the confinement effect on the glassy dynamics has also been discussed in a study of binary soft-disk mixtures with large size disparities~\cite{Weeks}. Voigtmann and Horbach~\cite{Voigtmann3} investigated the diffusion of a binary soft-sphere mixture with $\delta=0.35$. Emergence of the anomalous and power-law-like diffusion for the small particles has been interpreted as a precursor of a double-transition scenario, which combines a glass transition and a separate small-particle localization transition. This double-transition scenario was first identified using MCT for binary hard-sphere mixtures~\cite{Bosse1, Bosse2}. Very recently, it has been demonstrated by MCT calculations~\cite{Voigtmann4} that multiple glasses can occur even in the simplest size-asymmetric binary hard-sphere mixtures. These different kinds of glasses can be distinguished by considering whether small particles remain mobile and whether small particles contribute significantly to perturbing the big-particle structure, and can be separated by sharp transitions that give rise to higher-order transition phenomena involving logarithmic decay of the relaxation process and power-law-like behavior of diffusion. In particular, the glass transition curve displays an ``inverted shape'' for sufficiently large size disparity. Finally, it should be noted that highly asymmetric binary mixtures also attract a lot of interest with the emphasis on their phase behavior~\cite{Dijkstra1, Dijkstra2, Dijkstra3, Imhof1, Imhof2} and rheological properties~\cite{Seyboldt, Henrich, Kruger}.

In spite of the above interesting findings, the structure, the compressibility factor, and in particular, the dynamics in highly size-asymmetric binary liquids are not well understood. In this work, we study highly size-asymmetric binary hard-disk liquids ($\delta=0.2$ and $0.15$) by tuning the area fraction of the small disks and thus we can control the strength of the confinement effects for the small disks and hence the strength of the depletion effects for the large disks. We find that the static pair correlations of the large disks are only weakly perturbed by adding small disks. By contrast, the higher-order static correlations of the large disks, captured by the bond-orientation correlation functions, can be strongly affected. Accordingly, the static correlation length of the large disks deduced from the bond-orientation correlation functions first decreases significantly and then tends to reach a plateau as the area fraction of the small disks increases. The compressibility factor of the highly size-asymmetric binary hard-disk liquids shows non-monotonic change with the area fraction of the small disks, which can be rationalized by separating different contributions to it. We further find that adding small disks can influence dynamics of the system in quantitative and qualitative ways. For the large disks, the structural relaxation time exhibits monotonic increase with increasing the area fraction of the small disks at low and moderate area fractions of the large disks and crosses over into ``reentrant'' behavior at sufficiently high area fractions of the large disks. This ``reentrant'' behavior strongly resembles the reentrant glass transition in short-ranged attractive colloids~\cite{Poon, Sciortino3} (where two distinct kinds of glasses exist, dominated respectively by repulsion and attraction) and the inverted glass transition in binary hard spheres with large size disparity~\cite{Voigtmann4}. By tuning the area fraction of the small disks, relaxation process for the small disks shows concave-to-convex crossover and logarithmic decay behavior, as found in other binary mixtures with large size disparity~\cite{Moreno1, Moreno2}. Moreover, diffusion of both species is suppressed by adding small disks. Long-time diffusion for the small disks shows power-law-like behavior at sufficiently large $\phi_{s}$ values, which implies that a glass transition for the large disks and a localization transition for the small disks can occur in our system.

The article is organized as follows. In Sec. II the technical details of our work is described. In Sec. III we first present the results of the glass transition in the absence of the small disks and then discuss the influence of adding small disks on the structure of the large disks, the compressibility factor of the system and the dynamics of the highly size-asymmetric binary hard-disk liquids. Finally, we conclude the paper in Sec. V.

\section{Model and Methods}

We perform event-driven molecular dynamics simulations ($EDMD$)~\cite{Allen} of quasi-binary hard disks with large size disparity under periodic boundary conditions. The number of the large disks is fixed to be $N_{l}=1000$. To avoid crystallization, the diameters for the large disks are chosen equidistantly in the range $0.8-1.2$ with an interval of $0.01$, retaining the average diameter $<\sigma_{l}>=1$, then the size polydispersity of the large disks is $\Delta=\sqrt{(<\sigma_{l}^{2}>-<\sigma_{l}>^{2})}/<\sigma_{l}>=11.98\%$, where $<\cdot\cdot\cdot>$ is the average of the corresponding variable among all the large particles. We will show later that crystallization is indeed avoided and glassy dynamics develops on increasing density for hard-disk systems with such polydispersity. To model a binary mixture with sufficiently large size disparity, the size ratio $\delta=\sigma_{s}/<\sigma_{l}>$ is chosen to be $0.2$ (or $0.15$). The area fraction of the large disks $\phi_{l}=\pi \sum_{j=1}^{N_{l}}\sigma_{l, j}^{2}/4L^{2}$, where $L$ denotes the simulation box dimension, ranges from $0.76$ to $0.784$. By adjusting the small disk number $N_{s}$, the area fraction of the small disks $\phi_{s}=\pi N_{s}\sigma_{s}^{2}/4L^{2}$ is varied from $0$ to $0.06$, then the largest numbers of the small disks are $1974$ and $3509$ for $\delta=0.2$ and $0.15$ respectively. All the particles have the same mass $m$, and the temperature $T$ is irrelevant for athermal systems and we set it to be $T=1.0$. Length, time and pressure are reported in units of $<\sigma_{l}>$, $\sqrt{m<\sigma_{l}>^{2}/k_{B}T}$ and $k_{B}T/<\sigma_{l}>^{2}$, where $k_{B}$ is the Boltzmann's constant. To properly create the initial configuration for the binary hard disks with large size disparity, the desired state with specific $\phi_{l}$ and $\phi_{s}$ is generated by compressing a low-density liquid using the Lubachesky-Stillinger algorithm~\cite{LS1, LS2, Donev}. The starting configuration is then equilibrated for $10^{5}$ before production run for each state point and $8$ independent runs are performed to improve the statistics.

\section{Results and discussion}

In this section, we first present the results of the glass transition in the absence of small disks. Then we discuss the effects of adding much smaller disks on the structure of the large disks, the compressibility factor and the dynamics of the highly asymmetric binary hard-disk mixtures. It is expected that the static pair correlations of the large disks will not be perturbed significantly by the small disks due to the large size disparity. However, we will show in the following that adding very small disks will strongly affect the higher-order static correlations of the large disks, the compressibility factor of the system and the dynamics of both species.

\subsection{Glass transition in the absence of small disks}

\begin{figure}[!htbp]
 \centering
 \includegraphics[angle=0,width=0.5\textwidth]{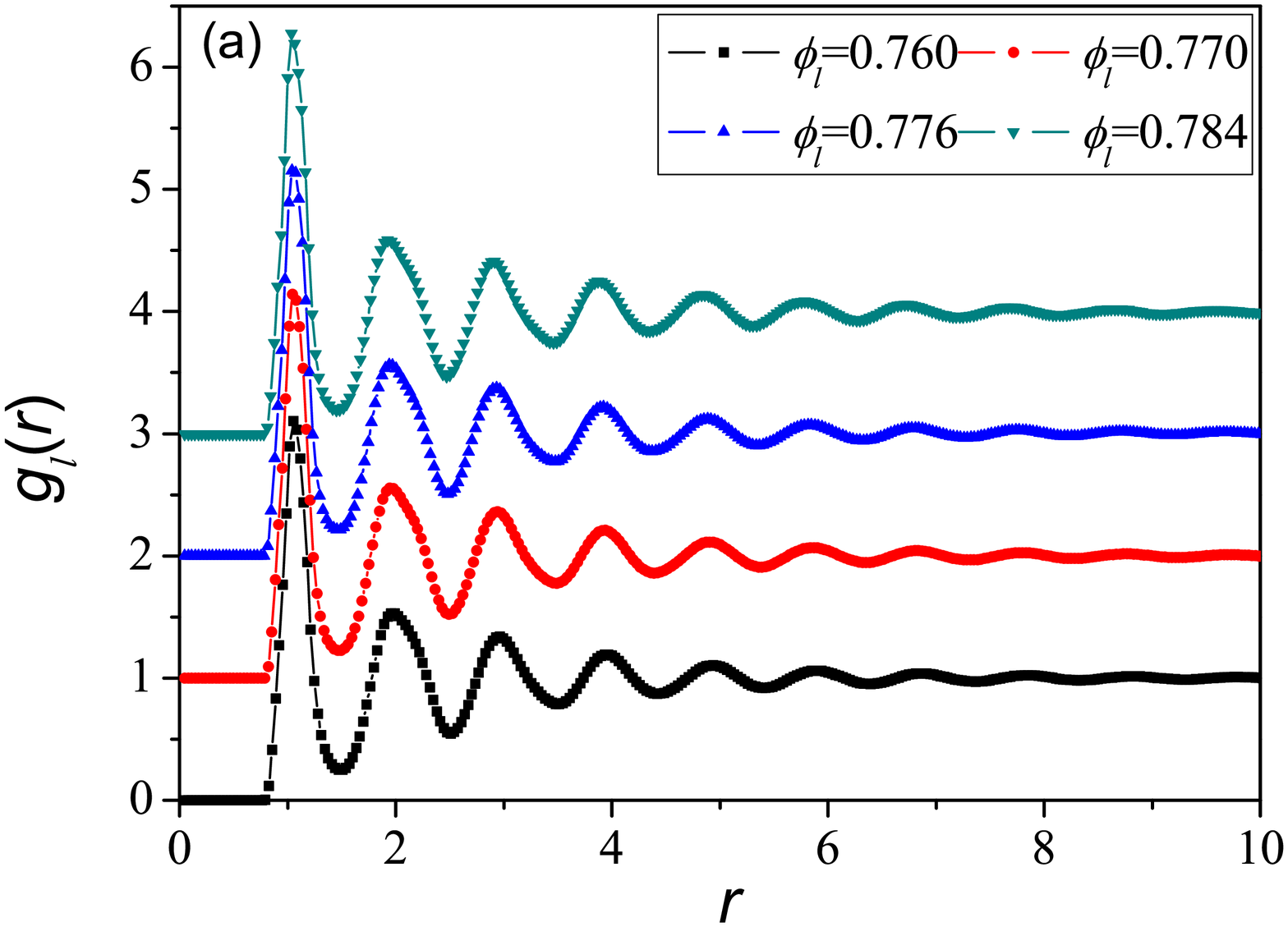}
 \includegraphics[angle=0,width=0.5\textwidth]{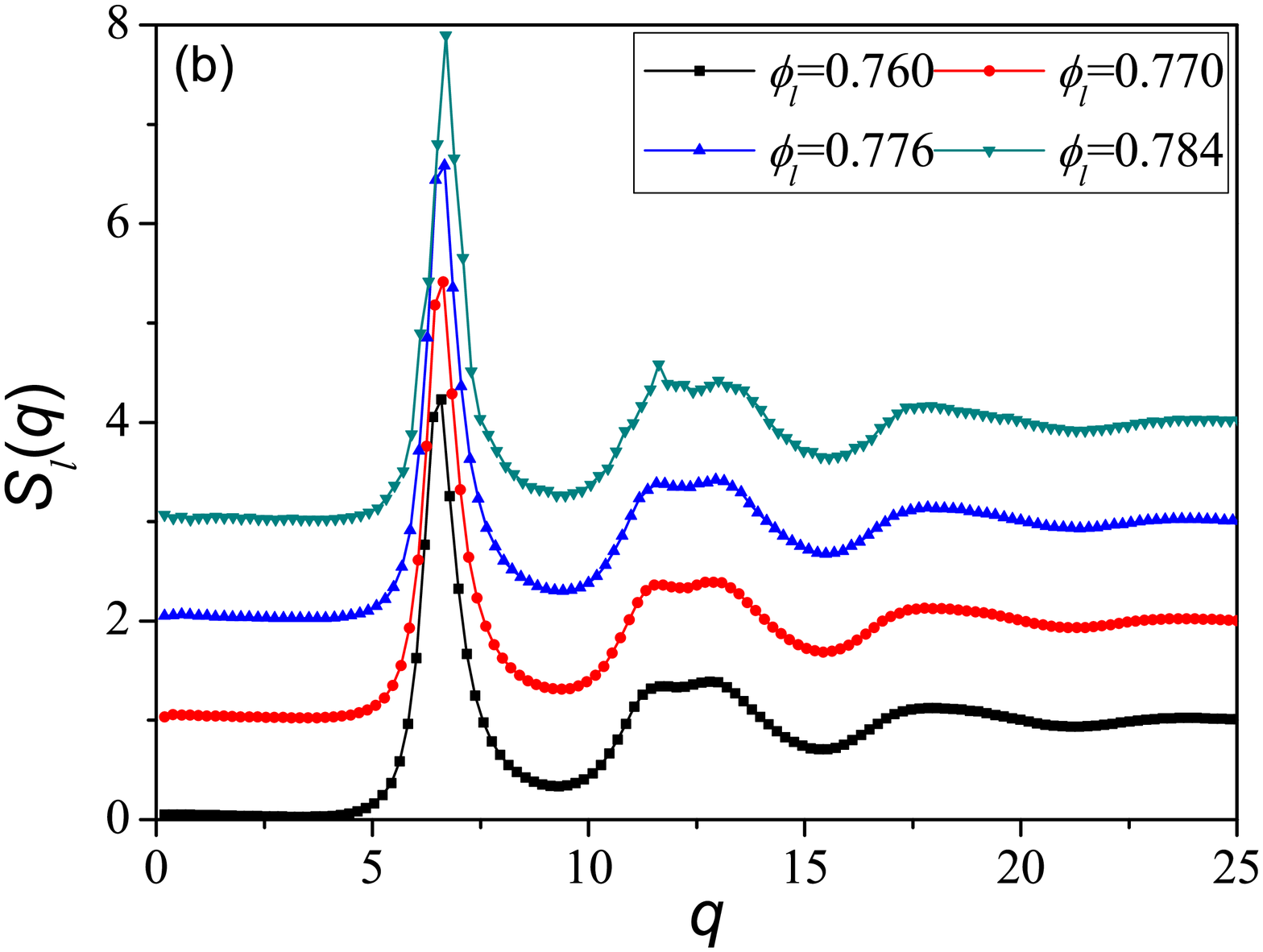}
 \caption{(a) The pair correlation function $g_{l}(r)$ and (b) the static structure factor $S_{l}(q)$ at low-$q$ region at varying $\phi_{l}$ for $\phi_{s}=0$.}
\end{figure}

\begin{figure}[!htbp]
 \centering
 \includegraphics[angle=0,width=0.5\textwidth]{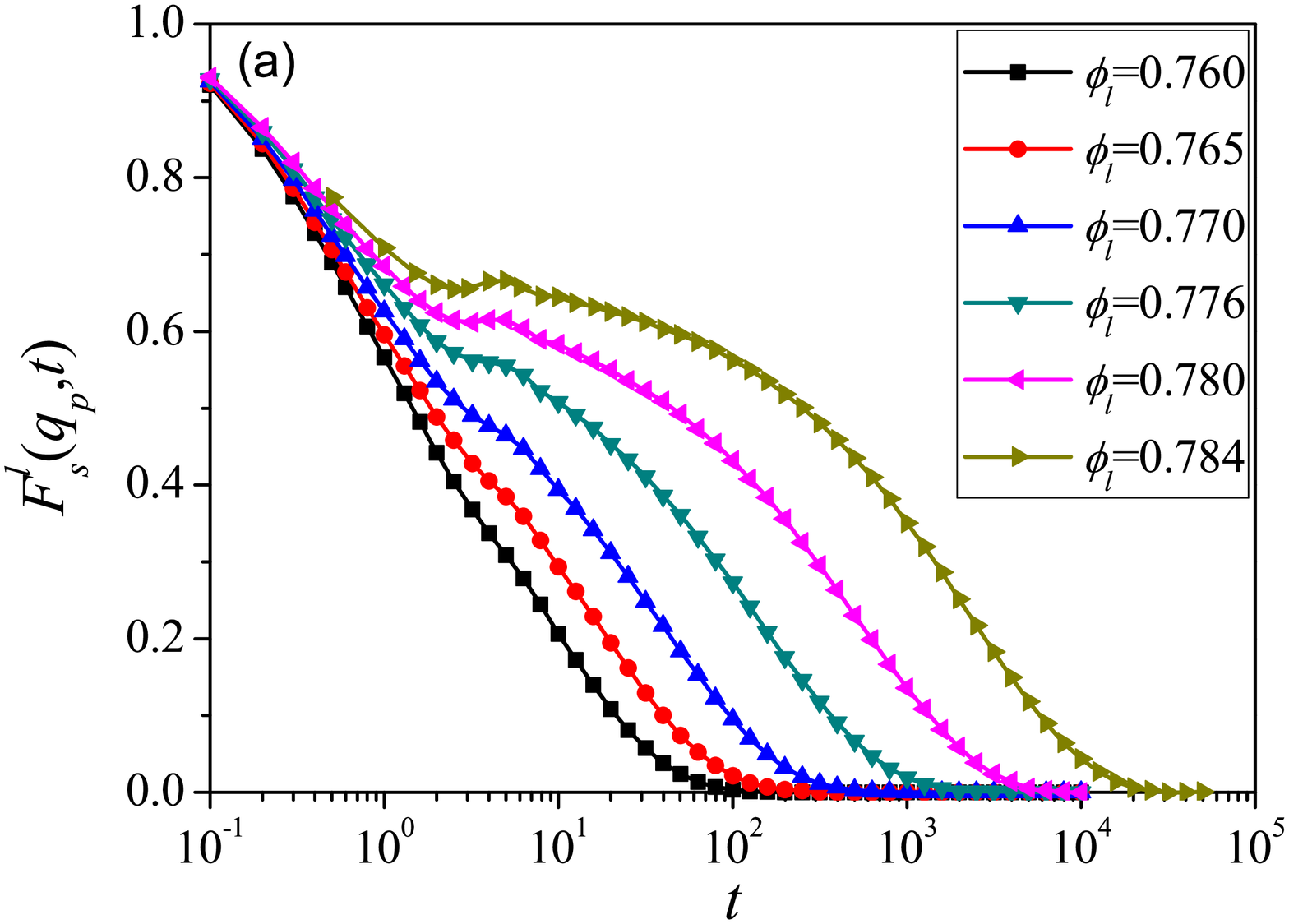}
 \includegraphics[angle=0,width=0.5\textwidth]{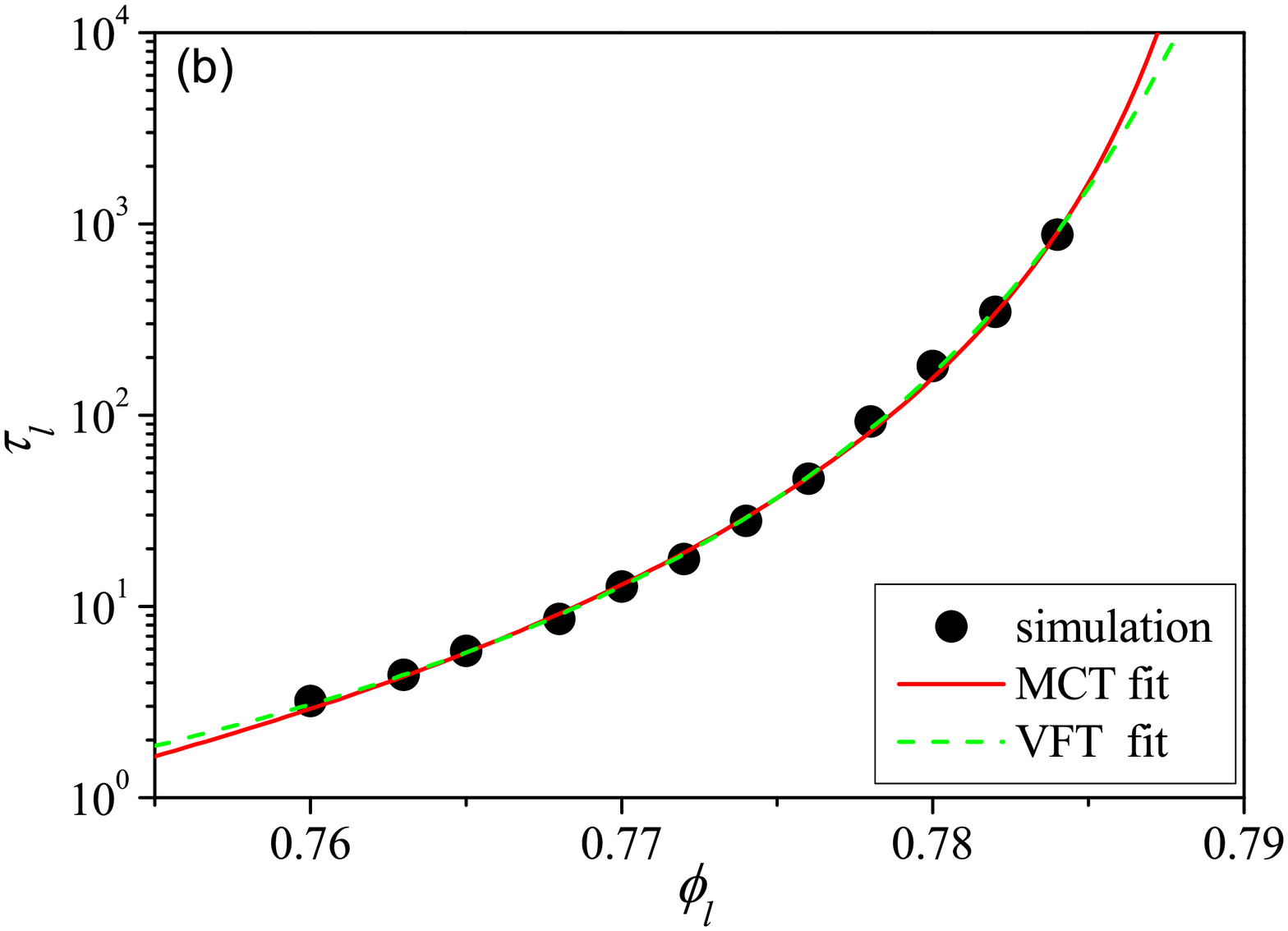}
 \caption{(a) The self-intermediate scattering function $F_{s}^{l}(q_{p},t)$ at varying $\phi_{l}$ for $\phi_{s}=0$. (b) $\phi_{l}$ dependence of $\alpha$-relaxation time $\tau_{\alpha}$ at $\phi_{s}=0$. The red solid line is the result of the MCT power-law fitting $\tau_{\alpha}\sim (\phi_{c}-\phi_{l})^{-\gamma}$ with $\gamma=3.8$ and $\phi_{c}=0.791$. The green dashed line is the result of the Vogel-Fulcher-Tamman fitting $\tau_{\alpha}\sim e^{D\phi_{l}/(\phi_{0}-\phi_{l})}$ with $D=0.28$ and $\phi_{c}=0.805$.}
\end{figure}

We first focus on the static structure of the liquids at $\phi_{s}=0$, as characterized by the pair correlation function and the static structure factor, which are defined as $g_{\alpha}(r)=\frac{L^2}{2\pi r\Delta rN_{\alpha}(N_{\alpha}-1)}\Sigma_{j \neq k}\delta(r-|\textbf{r}_{jk}|)$ and $S_{\alpha}(q)=\frac{1}{N_{\alpha}}<\rho_{\alpha}(q)\rho_{\alpha}(-q)>$, respectively. Here, $\alpha$ denotes the particle type ($l$ for large disks and $s$ for small disks), and $\rho_{\alpha}(q)=\sum_{j=1}^{N_{\alpha}}e^{i\textbf{q} \cdot \textbf{r}_{j}}$ with $\textbf{r}_{j}$ the position of particle $j$. The representative results are displayed in Fig. 1. It is seen that both $g_{l}(r)$ and $S_{l}(q)$ vary weakly when $\phi_{l}$ is increased from $0.76$ to $0.784$. However, the dynamics can slow down by orders of magnitude in the same density range, which will be shown later. Additionally, although crystallization is obviously prevented, the splitting second peaks in $g_{l}(r)$ and $S_{l}(q)$ become more apparent as $\phi_{l}$ increases (this phenomenon is more apparent in the results of $S_{l}(q)$ for our model), suggesting the development of the locally preferred order on increasing density (i.e., the system is weakly frustrated and the local order is indeed hexatic in two dimensions). These static structural features on approaching the glass transition have been discussed extensively in the past few years~\cite{Xu}. Evidence has been provided that the development of the locally preferred order can be connected to slow dynamics and dynamic heterogeneity of glass-forming liquids.

The dynamic slowing down on approaching the glass transition can be characterized by the self-intermediate scattering function $F_{s}^{\alpha}(q_{p},t)=\frac{1}{N_{\alpha}}<\sum_{j=1}^{N_{\alpha}}e^{i\textbf{q}_{p}\cdot[\textbf{r}_{j}(t)-\textbf{r}_{j}(0)]}>$, where $i=\sqrt{-1}$ and the wave number $q_{p}$ corresponds to the first peak of the static structure factor for the large disks. Fig. 2(a) shows the results for several densities. We observe that the relaxation slows down and becomes more stretched as $\phi_{l}$ increases. In particular, a two-step decay emerges at sufficiently large $\phi_{l}$ values. This two-step process is a typical phenomenon in glass-forming liquids, and it reflects the increasingly caged motion of particles at high area fractions. We define the structural relaxation time $\tau_{\alpha}$ as $F_{s}^{\alpha}(q_{p},t=\tau_{\alpha})=1/e$. In Fig. 2(b) we plot $\tau_{l}$ as a function of $\phi_{l}$. Clearly, $\tau_{l}$ drastically increases as $\phi_{l}$ increases. For the density range investigated, the simulation data can be well fitted by the MCT power-law: $\tau_{l}\sim (\phi_{c}-\phi_{l})^{-\gamma}$, where $\phi_{c}$ is the MCT glass transition point, or the Vogel-Fulcher-Tamman (VFT) law: $\tau_{l}\sim e^{D\phi_{l}/(\phi_{0}-\phi_{l})}$, where $D$ is the fragility index and $\phi_{0}$ is the ideal glass-transition point. MCT fitting yields $\gamma=3.8$ and $\phi_{c}=0.791$ and VFT fitting yields $D=0.28$ and $\phi_{0}=0.805$.

Therefore, we have demonstrated that the model presented here possesses the ability to undergo the liquid-glass transition and reproduces essential aspects of a typical glass-forming liquid. In the following sections, we will discuss the influence of the small disks on the structure of the large disks, the compressibility factor and the dynamics of the binary mixtures.

\subsection{Influence of small disks on the structure of the large disks}

\begin{figure}[!htbp]
 \centering
 \includegraphics[angle=0,width=0.5\textwidth]{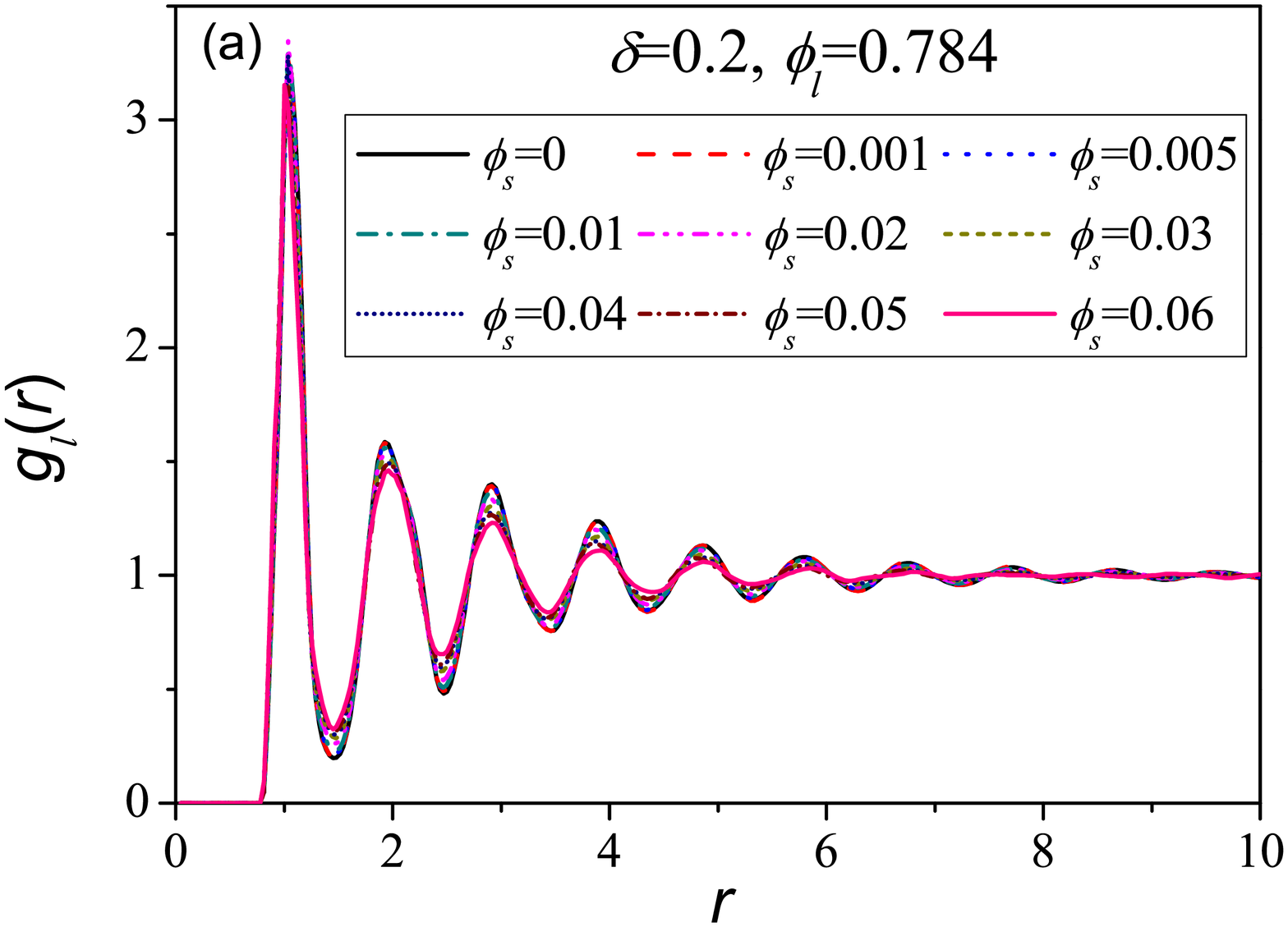}
 \includegraphics[angle=0,width=0.5\textwidth]{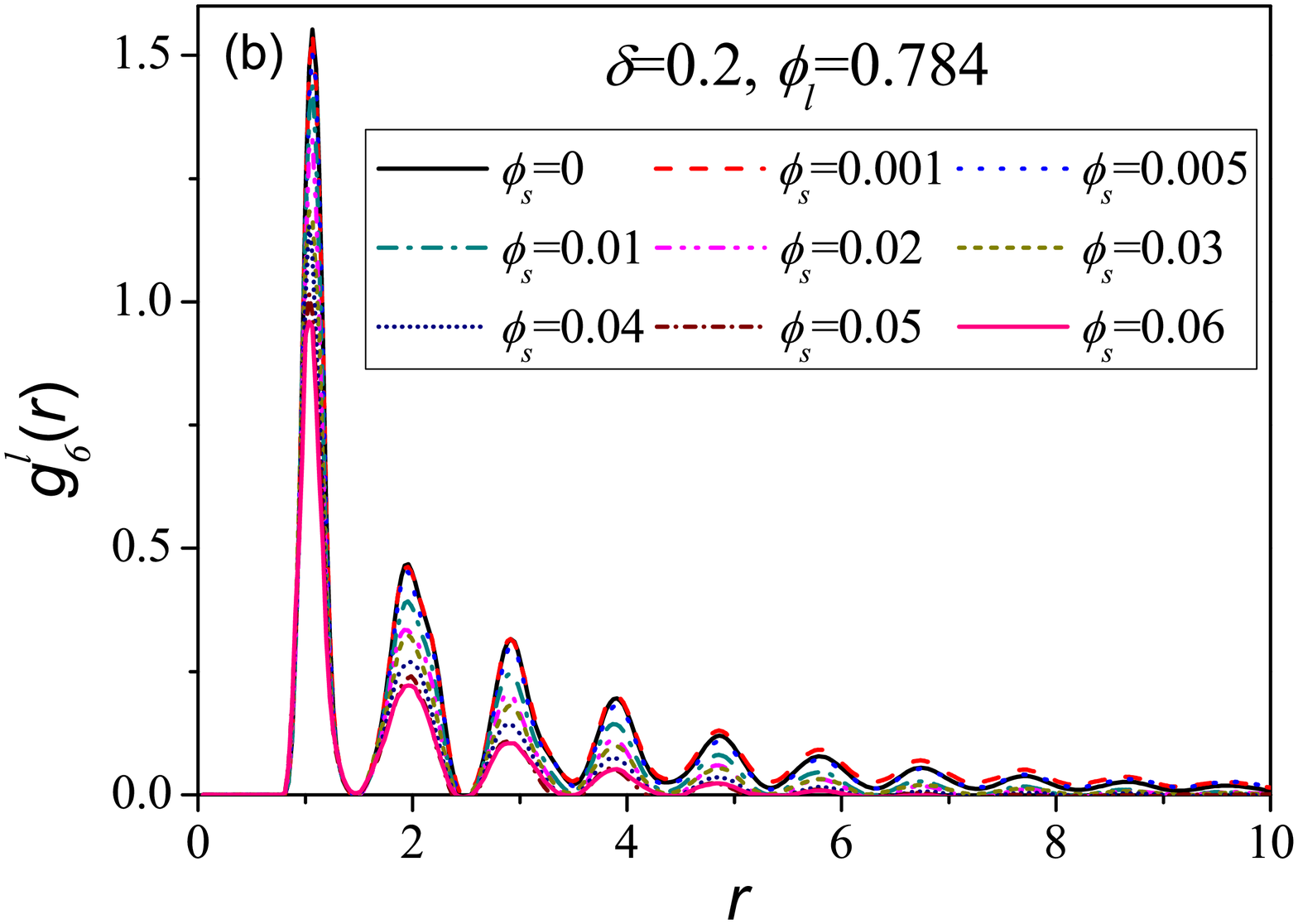}
 \caption{(a) The pair correlation function $g_{l}(r)$ and (b) the bond-orientation correlation function $g_{6}(r)$ at varying $\phi_{s}$ for $\delta=0.2$ and $\phi_{l}=0.784$.}
\end{figure}

\begin{figure}[!htbp]
 \centering
 \includegraphics[angle=0,width=0.5\textwidth]{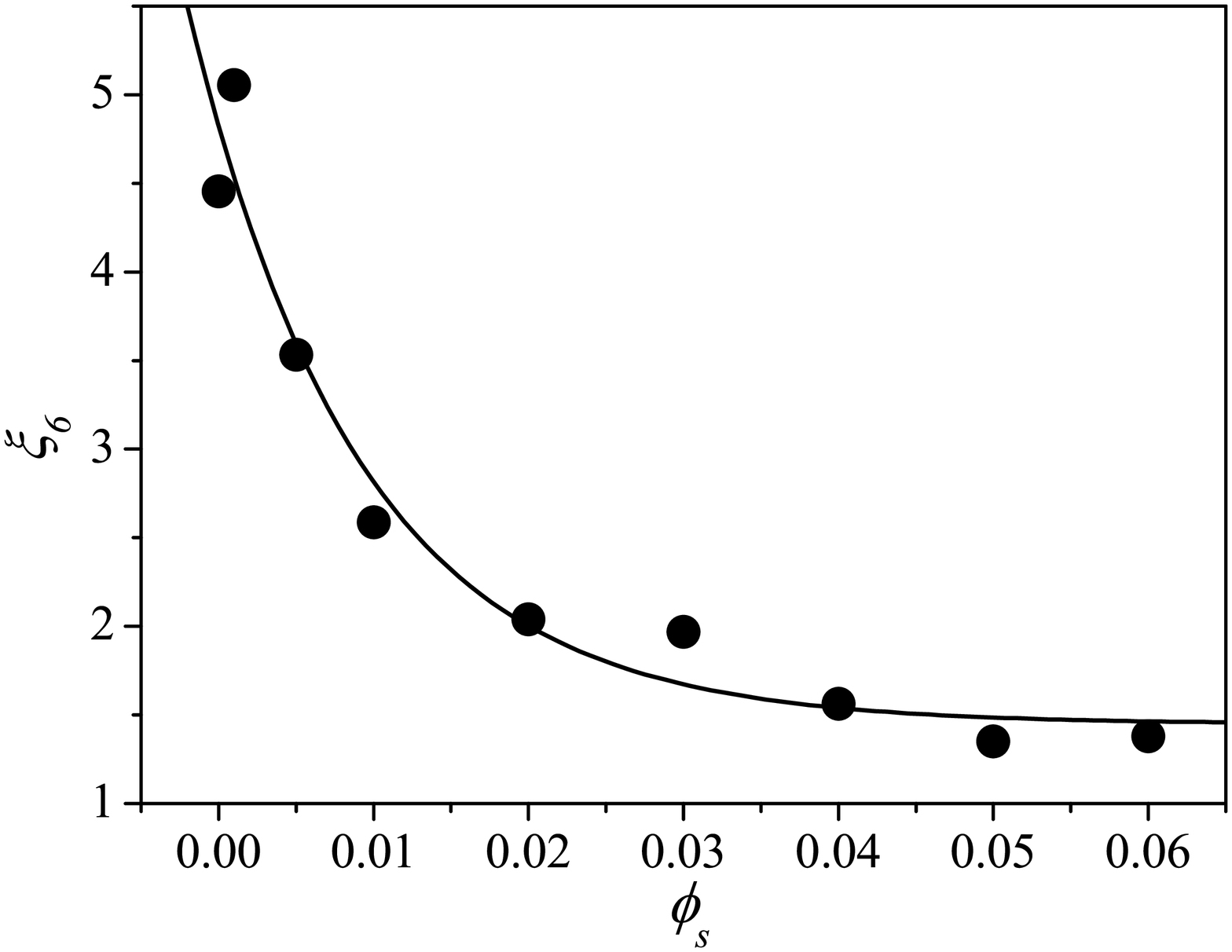}
 \caption{$\phi_{s}$ dependence of static correlation length $\xi_{6}$ at $\delta=0.2$ and $\phi_{l}=0.784$. $\xi_{6}$ is obtained by fitting $g_{6}(r)/g(r)$ with the OZ function (see text). The solid line is guide for the eyes.}
\end{figure}

Since the size disparity in this work is very large, it is expected that addition of the small disks will not perturb the static correlations significantly. This is indeed seen in $g_{l}(r)$ (Fig. 3(a)) and $S_{l}(q)$ (data not shown) at varying $\phi_{s}$. Small but noticeable differences are also exhibited in Fig. 3(a): The first minimum of $g_{l}(r)$ slightly increases as $\phi_{s}$ increases and $g_{l}(r)$ on the whole changes mildly with $\phi_{s}$. This observation is similar to the results of Ref.~\cite{Moreno2} for soft-sphere mixtures. Then it is interesting to assess whether the higher-order static correlations are strongly influenced by adding small disks since the importance of higher-order static correlations, in understanding the glass transition, has been emphasized in recent years~\cite{Tanaka5, Sausset2, Tarjus, Coslovich}. Because our system can form some locally preferred ordered structures (already indicated in Fig .1), we can use the bond-orientation correlation function $g_{6}^{l}(r)=\frac{L^2}{2\pi r\Delta rN_{l}(N_{l}-1)}\Sigma_{j\neq k}\delta(r-|\textbf{r}_{jk}|)\psi_{6}^{j}\psi_{6}^{k*}$ to characterize static correlations beyond the pair level. Here, $\psi_{6}^{j}$ is a sixfold bond-orientation order parameter, and defined as $\psi_{6}^{j}=\frac{1}{n_{j}}\sum_{m=1}^{n_{j}}e^{i6\theta_{m}^{j}}$, where $n_{j}$ is the number of the nearest neighbors for particle $j$ and determined by the Voronoi construction~\cite{Allen}, and $\theta_{m}^{j}$ is the angle between $(\textbf{r}_{m}-\textbf{r}_{j})$ and the $x$ axis (particle $m$ is a neighbor of particle $j$). These higher-order correlations reveal more clearly the structural change of the large disks as $\phi_{s}$ increases, as shown in Fig. 3(b). The envelops of $g_{6}^{l}(r)/g_{l}(r)$ can be well fitted by the Ornstein-Zernike (OZ) function $r^{-1/2}e^{-r/\xi_{6}}$, and the obtained fitting parameters $\xi_{6}$ stand for the static correlation lengths, which are shown in Fig. 4. It is found that $\xi_{6}$ first dramatically decreases as $\phi_{s}$ increases and then tends to reach a plateau at $\phi_{s}>0.04$. Therefore, although the addition of the small disks does not perturb the structure of the glass-forming liquids at the static pair level, the higher-order static correlations can be strongly influenced. Since we will further show that the dynamics of the large disks can be strongly affected by adding small disks, our results also imply the important role of higher-order static correlations in understanding the dynamics of glass-forming liquids.

\subsection{Compressibility factor of highly asymmetric binary hard disks}

\begin{figure}[!htbp]
 \centering
 \includegraphics[angle=0,width=0.5\textwidth]{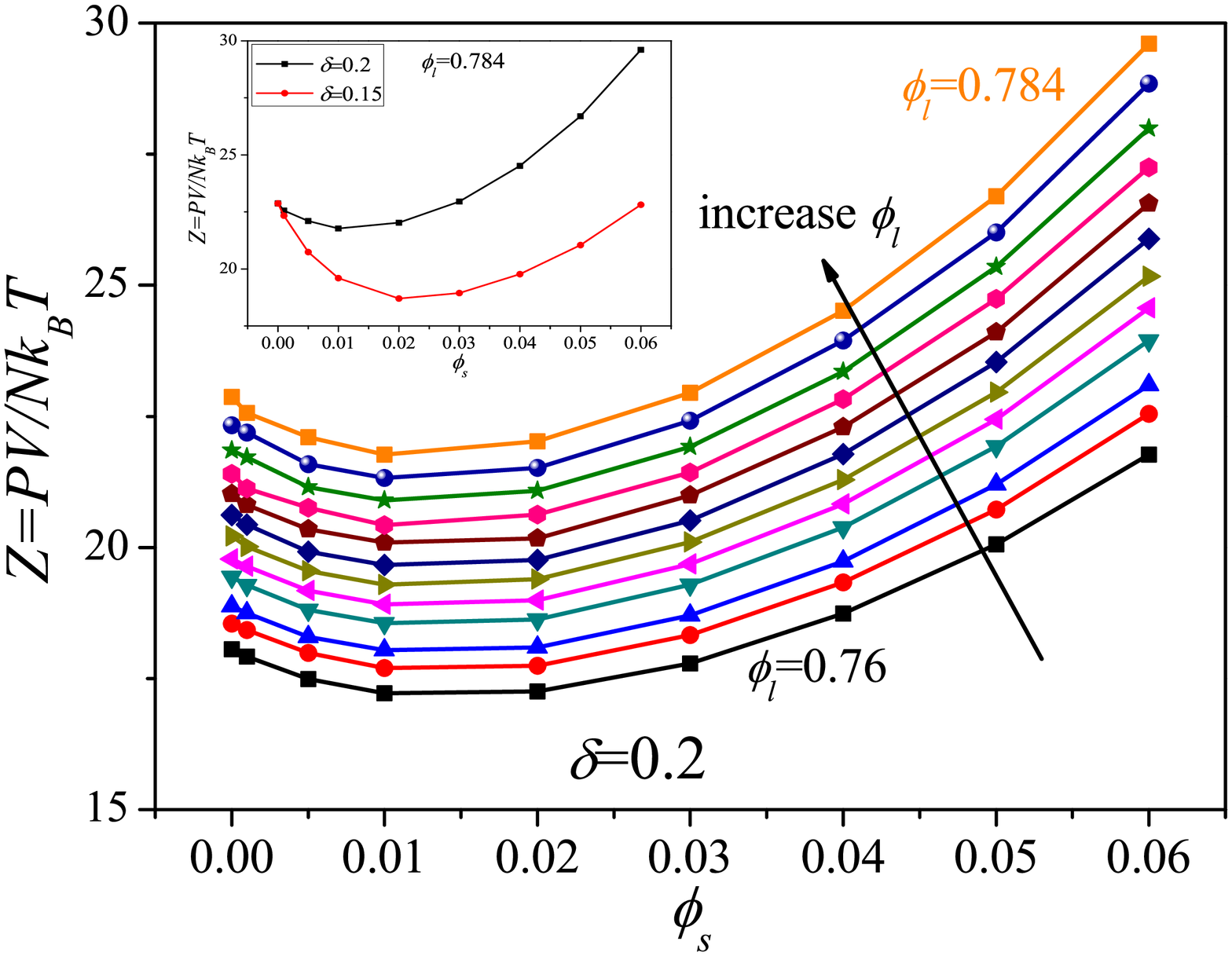}
 \caption{Main: $\phi_{s}$ dependence of the compressibility factor $Z$ at varying $\phi_{l}$ for $\delta=0.2$. Inset: effect of $\delta$ on $Z$ at $\phi_{l}=0.784$.}
\end{figure}

\begin{figure}[!htbp]
 \centering
 \includegraphics[angle=0,width=0.5\textwidth]{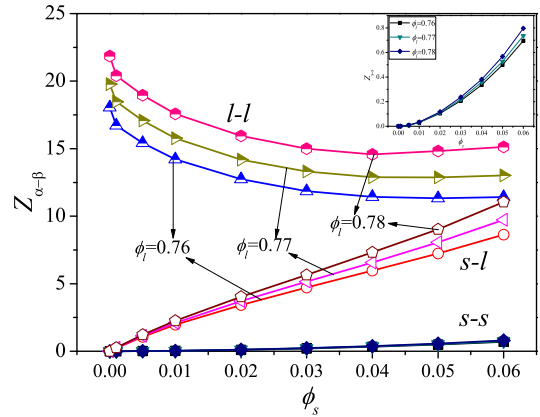}
 \caption{Different contributions to the total $Z$ for $\delta=0.2$ at several $\phi_{l}$ values. s-s, s-l and l-l indicate contributions from small-small, small-large and large-large particle collisions. The zoom in the inset highlights the contribution from the small-small particle collision. The results are similar for $\delta=0.15$.}
\end{figure}

We now focus on the thermodynamics of the highly asymmetric binary hard-disk mixtures. For hard-particle systems, the only relevant thermodynamic quantity is the pressure $P$. The compressibility factor $Z$ (and hence $P$) can be measured in an $EDMD$ by total momentum exchange from all the inter-particle collisions during a certain time interval $\Delta t$, i.e., $Z=\frac{PV}{Nk_{B}T}=1+\sum_{\Delta t}\frac{\Delta \textbf{r}_{jk}\cdot \Delta \textbf{v}_{j}}{2NT\Delta t}$ with $N$ the total particle number. It is found that as $\phi_{s}$ increases, $P$ (data not shown) monotonically increases due to the increase of the total particle number (since $N_{l}$ is fixed). However, the evolution of $Z$ with $\phi_{s}$ is not monotonic, as shown in Fig. 5. As $\phi_{s}$ increases, $Z$ first decreases and then increases, showing a minimum at some $\phi_{s}^{m}$, which shifts to larger values as the size disparity enhances ($\phi_{s}^{m}\approx0.01$ for $\delta=0.2$ and $\phi_{s}^{m}\approx0.02$ for $\delta=0.15$), as evidenced in the inset of Fig. 5. Additionally, $Z$ decreases at fixed $\phi_{l}$ and $\phi_{s}$ values as $\delta$ decreases.

To better understand the non-monotonic change of $Z$ with $\phi_{s}$, we can calculate different contributions to $Z$. Since in binary hard-particle systems, the following relation holds: $Z=1+Z_{s-s}+Z_{s-l}+Z_{l-l}$, where $Z_{s-s}$, $Z_{s-l}$ and $Z_{l-l}$ indicate contributions from small-small, small-large and large-large particle collisions and $Z_{\alpha-\beta}=\sum_{\Delta t}\frac{\Delta \textbf{r}_{\alpha\beta}\cdot \Delta \textbf{v}_{\alpha}}{2NT\Delta t}$. The results are shown in Fig. 6. We observe that as $\phi_{s}$ increases, $Z_{l-l}$ first decreases significantly and then tends to reach a plateau at $\phi_{s}\approx 0.04$, implying that the large-large particle collision frequency becomes low as $\phi_{s}$ increases and reaches nearly constant values at large $\phi_{s}$ values. Interestingly, the value of $\phi_{s}$ where $Z_{l-l}$ approximately ceases to decrease is consistent with that where $\xi_{6}$ roughly ceases to decrease. By contrast, both $Z_{s-l}$ and $Z_{s-s}$ increase rapidly within the whole $\phi_{s}$ range. However, their absolute values are small as compared to $Z_{l-l}$ due to the very large size disparity. Therefore, when $\phi_{s}$ is small, the amount of the increase from $Z_{s-l}$ and $Z_{s-s}$ cannot compensate the amount of the decrease from $Z_{l-l}$, resulting into the first decrease of $Z$. As $\phi_{s}$ increases further, the decrease of $Z_{l-l}$ becomes mild. In this case, however, binary collisions occur more frequently between small and large particles and between small and small particles and the other two contributions ($Z_{s-l}$ and $Z_{s-s}$) still increase rapidly, which leads to the increase of $Z$ at large $\phi_{s}$ values. Moreover, as can be seen in the inset of Fig .6, the increase rate of $Z_{s-s}$ becomes large as $\phi_{s}$ increases, implying enhancement of bulk-like effects for the small disks. We will show in the next that these effects can influence the dynamics of both species in quantitative and qualitative ways.

\subsection{Dynamics of highly asymmetric binary hard disks}

\begin{figure}[!htbp]
 \centering
 \includegraphics[angle=0,width=0.48\textwidth]{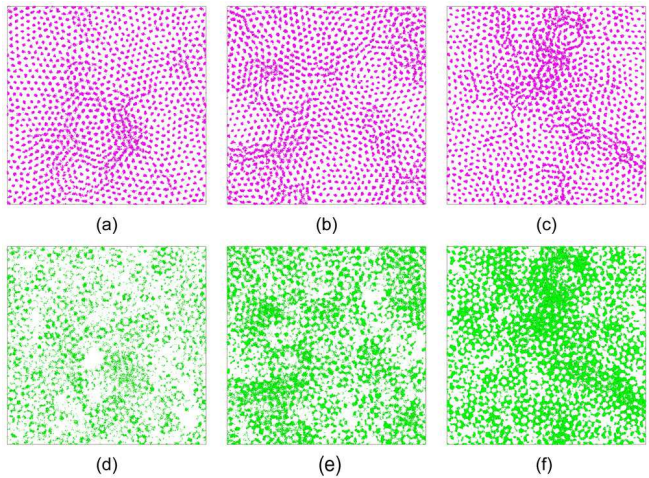}
 \caption{Particle trajectories during a time interval of $\tau_{l}$ at $\delta=0.2$ and $\phi_{l}=0.784$ for $\phi_{s}=0.01$ ((a), (d)), $\phi_{s}=0.03$ ((b), (e)) and $\phi_{s}=0.05$ ((c), (f)). (a), (b), (c) are for large disks and (d), (e), (f)) for small disks.}
\end{figure}

\begin{figure}[!htbp]
 \centering
 \includegraphics[angle=0,width=0.5\textwidth]{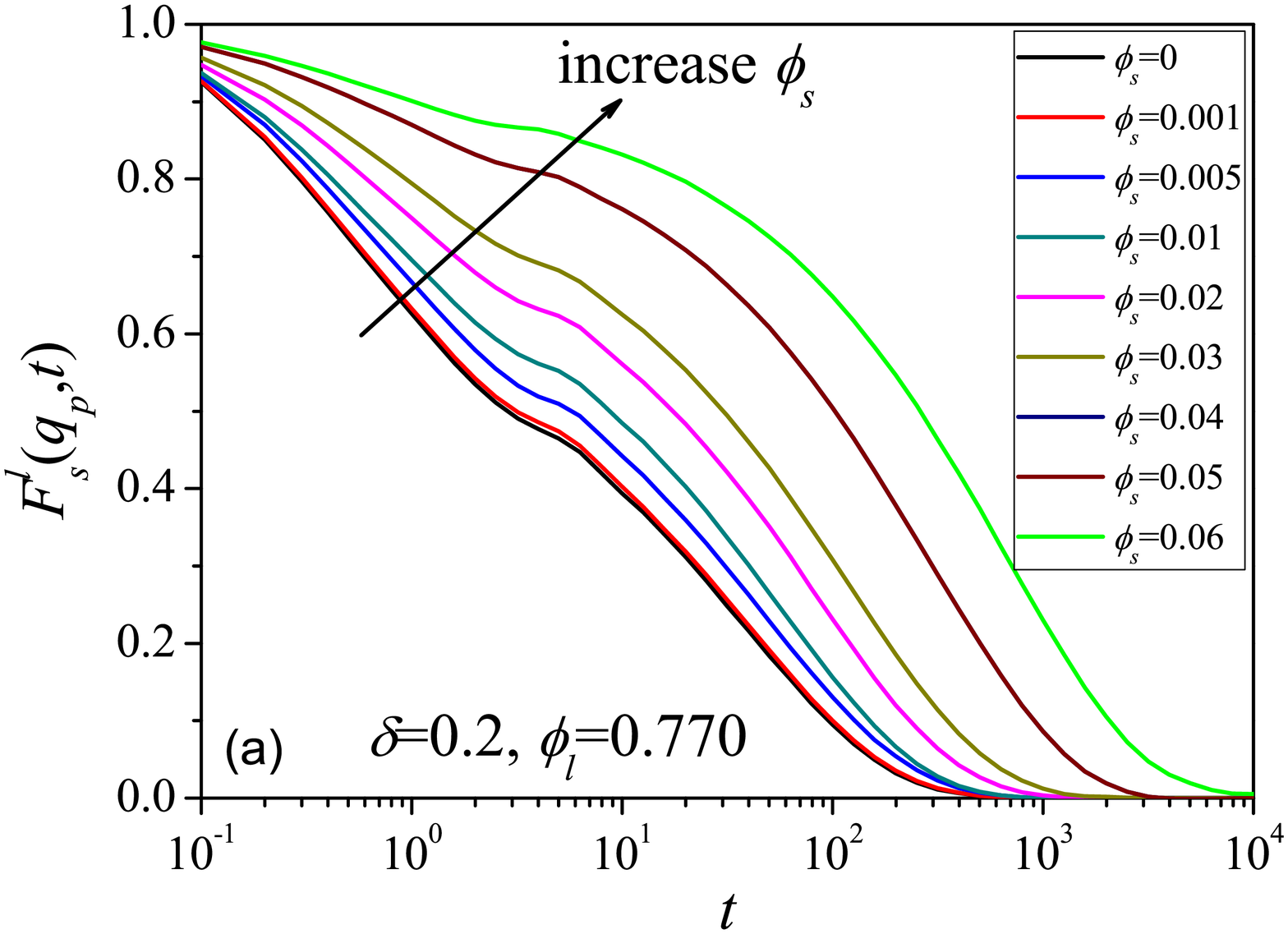}
 \includegraphics[angle=0,width=0.5\textwidth]{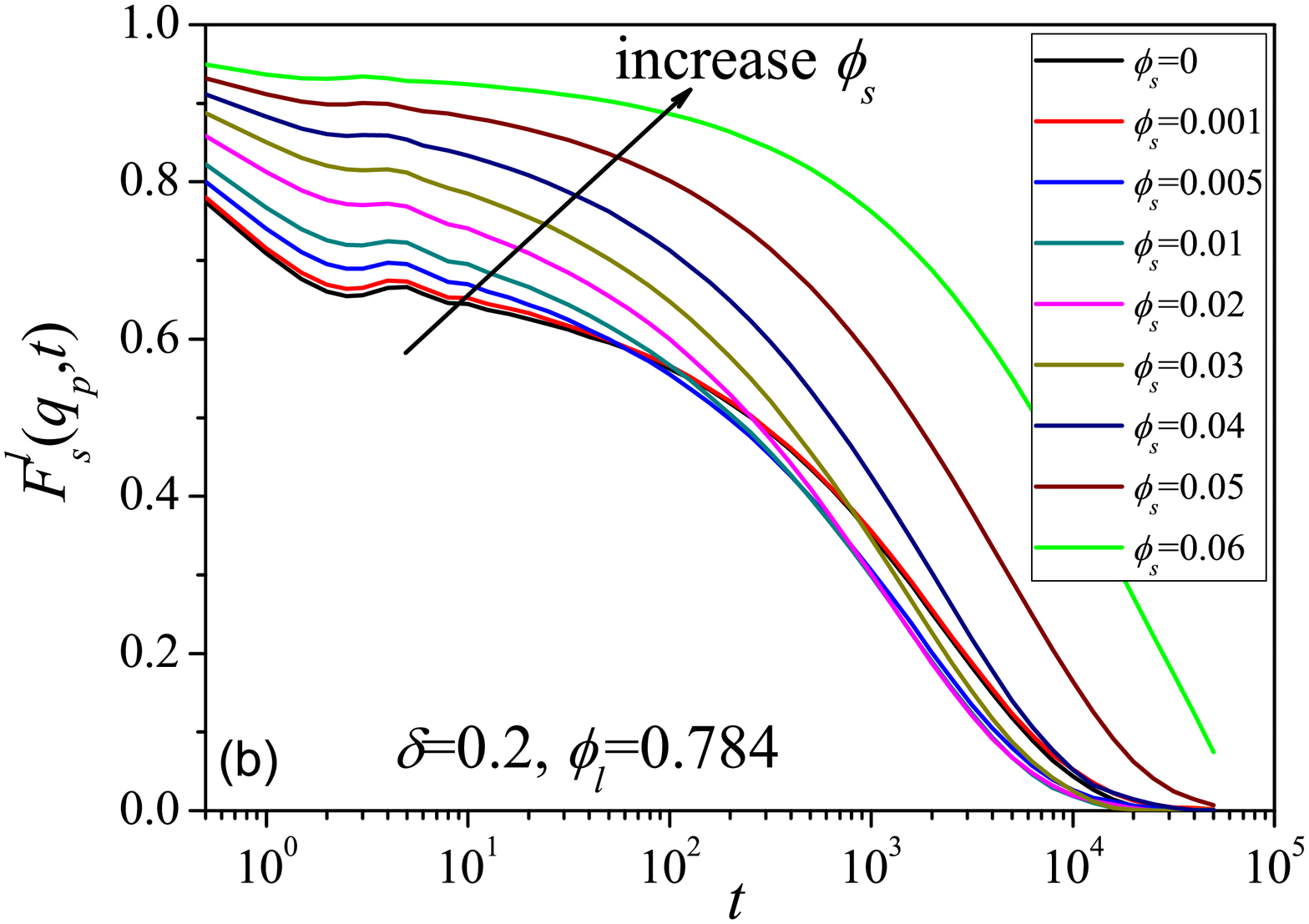}
 \caption{$\phi_{s}$ evolution of the self-intermediate scattering functions of the large disks $F_{s}^{l}(q_{p},t)$ at $\delta=0.2$ for (a) $\phi_{l}=0.77$ and (b) $\phi_{l}=0.784$. Here, $q_{p}$ corresponds to the first peak of $S_{l}(q)$ for $\phi_{s}=0$. The results are similar for $\delta=0.15$.}
\end{figure}

\begin{figure}[!htbp]
 \centering
 \includegraphics[angle=0,width=0.5\textwidth]{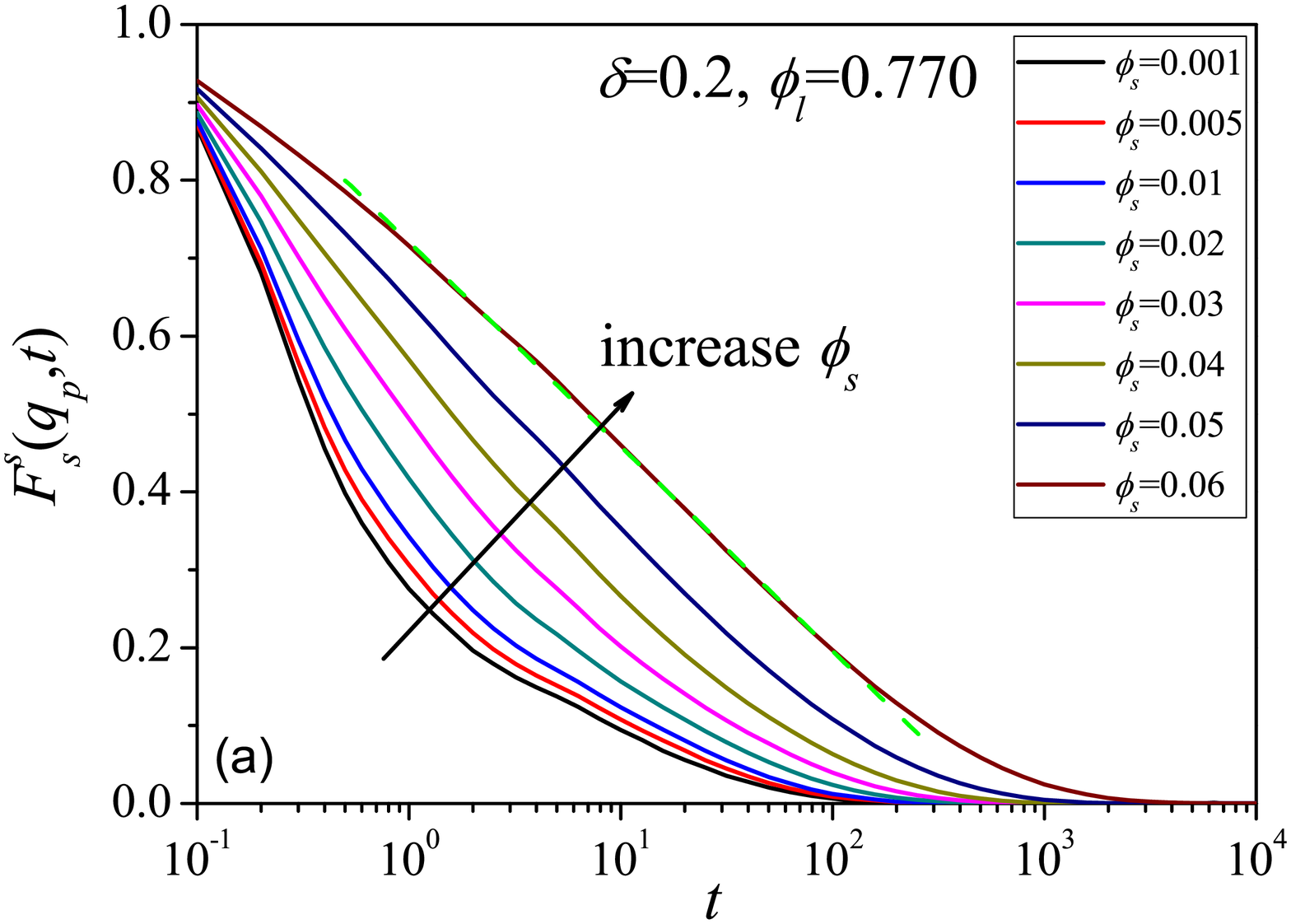}
 \includegraphics[angle=0,width=0.5\textwidth]{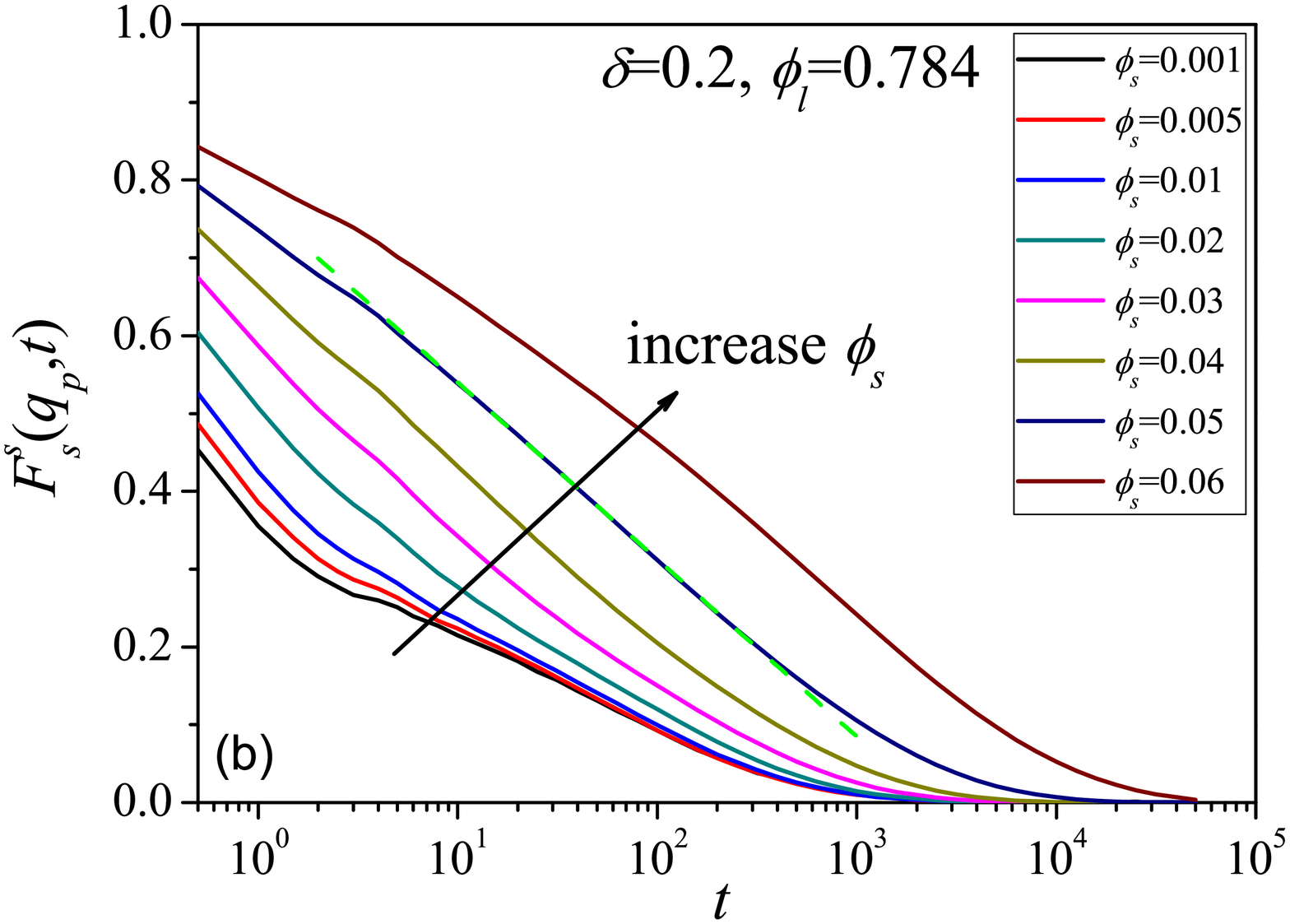}
 \caption{$\phi_{s}$ evolution of the self-intermediate scattering functions of the small diks $F_{s}^{s}(q_{p},t)$ at $\delta=0.2$ for (a) $\phi_{l}=0.77$ and (b) $\phi_{l}=0.784$. Here, $q_{p}$ corresponds to the first peak of $S_{l}(q)$ for $\phi_{s}=0$. The green dashed lines highlight the logarithmic decay of $F_{s}^{s}(q_{p},t)$. The results are similar for $\delta=0.15$.}
\end{figure}

\begin{figure}[!htbp]
 \centering
 \includegraphics[angle=0,width=0.5\textwidth]{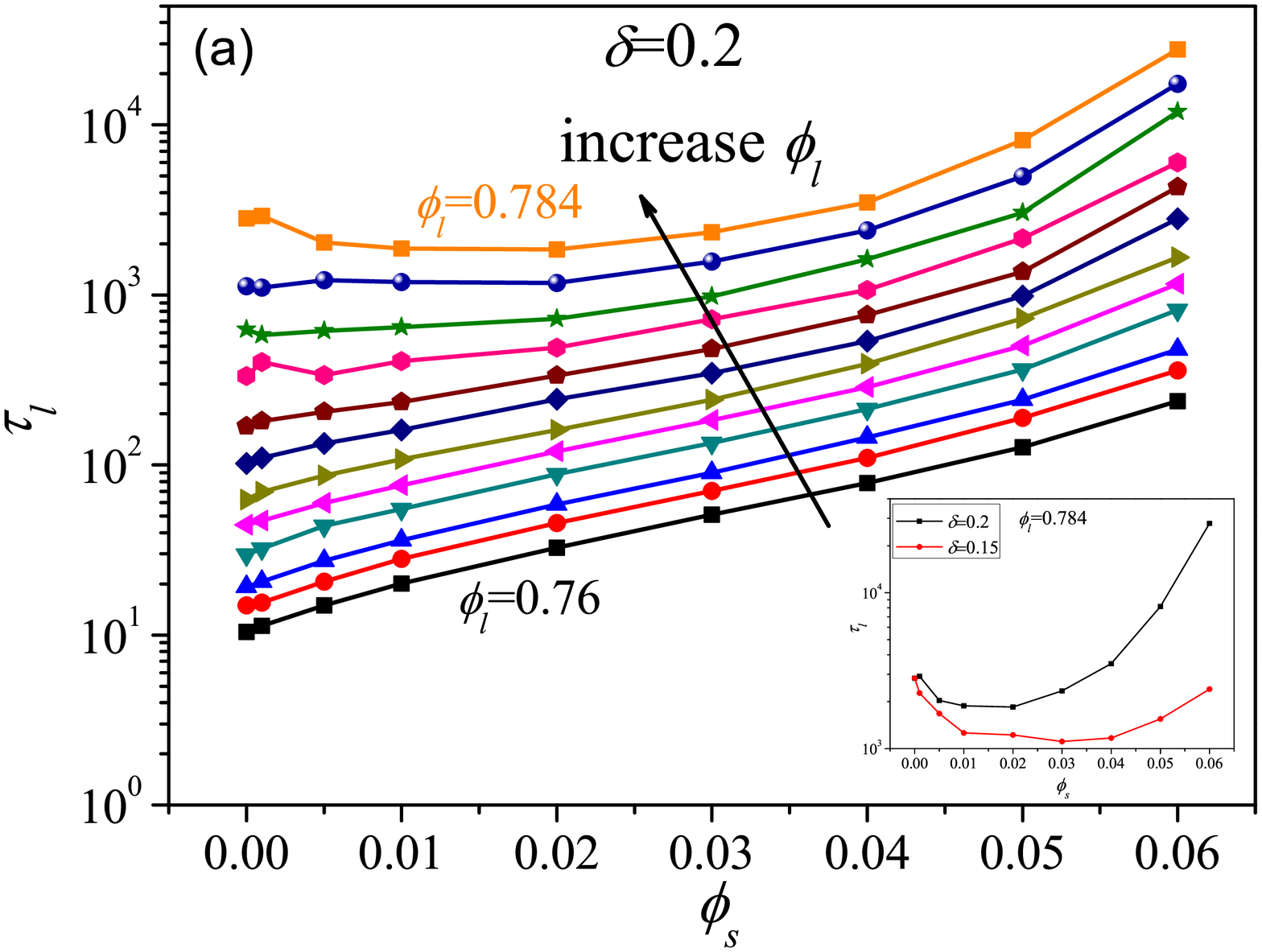}
 \includegraphics[angle=0,width=0.5\textwidth]{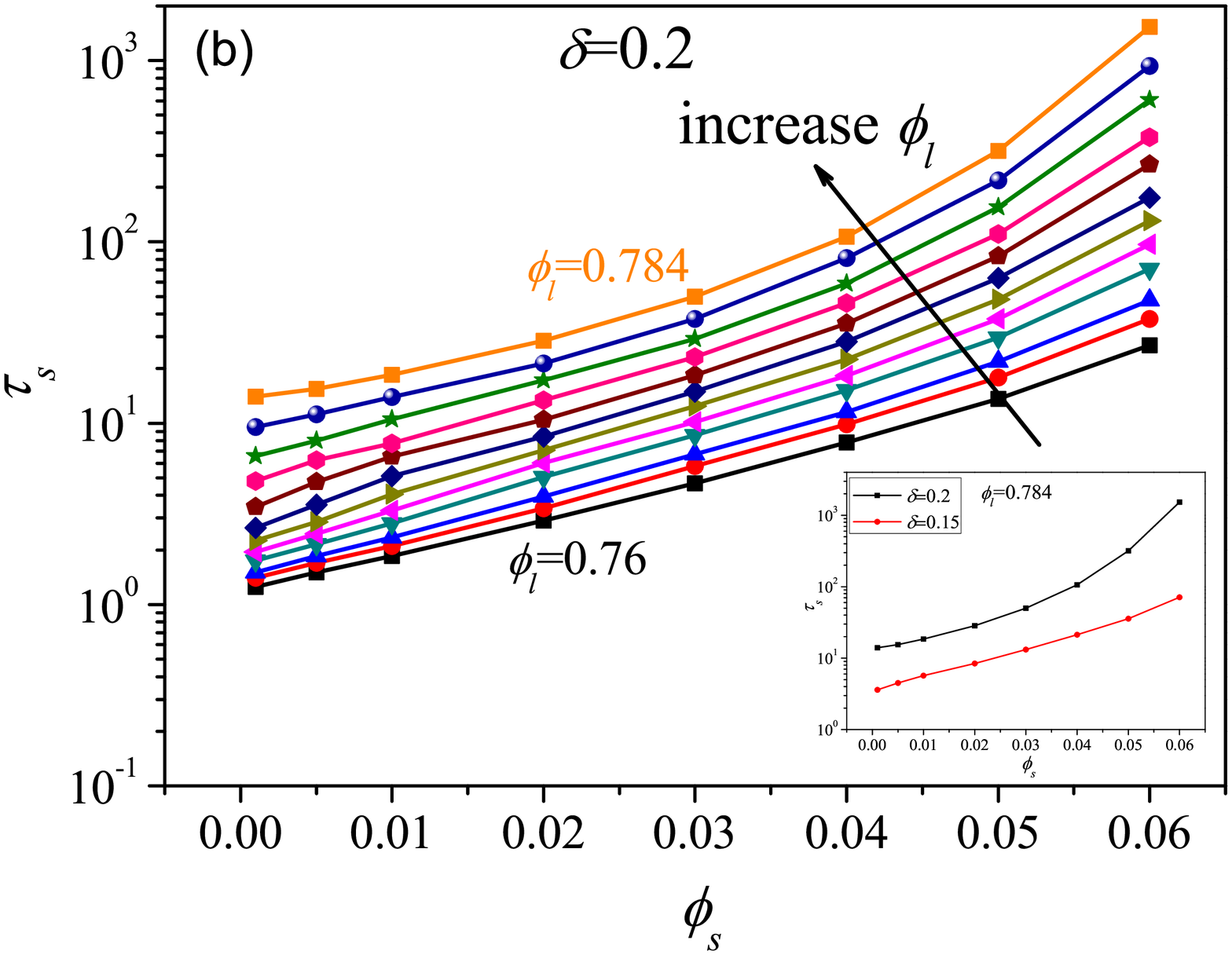}
 \caption{Main: $\phi_{s}$ evolution of relaxation times at $\delta=0.2$ for (a) large and (b) small disks. Inset: Effect of $\delta$ on relaxation times.}
\end{figure}

\begin{figure}[!htbp]
 \centering
 \includegraphics[angle=0,width=0.5\textwidth]{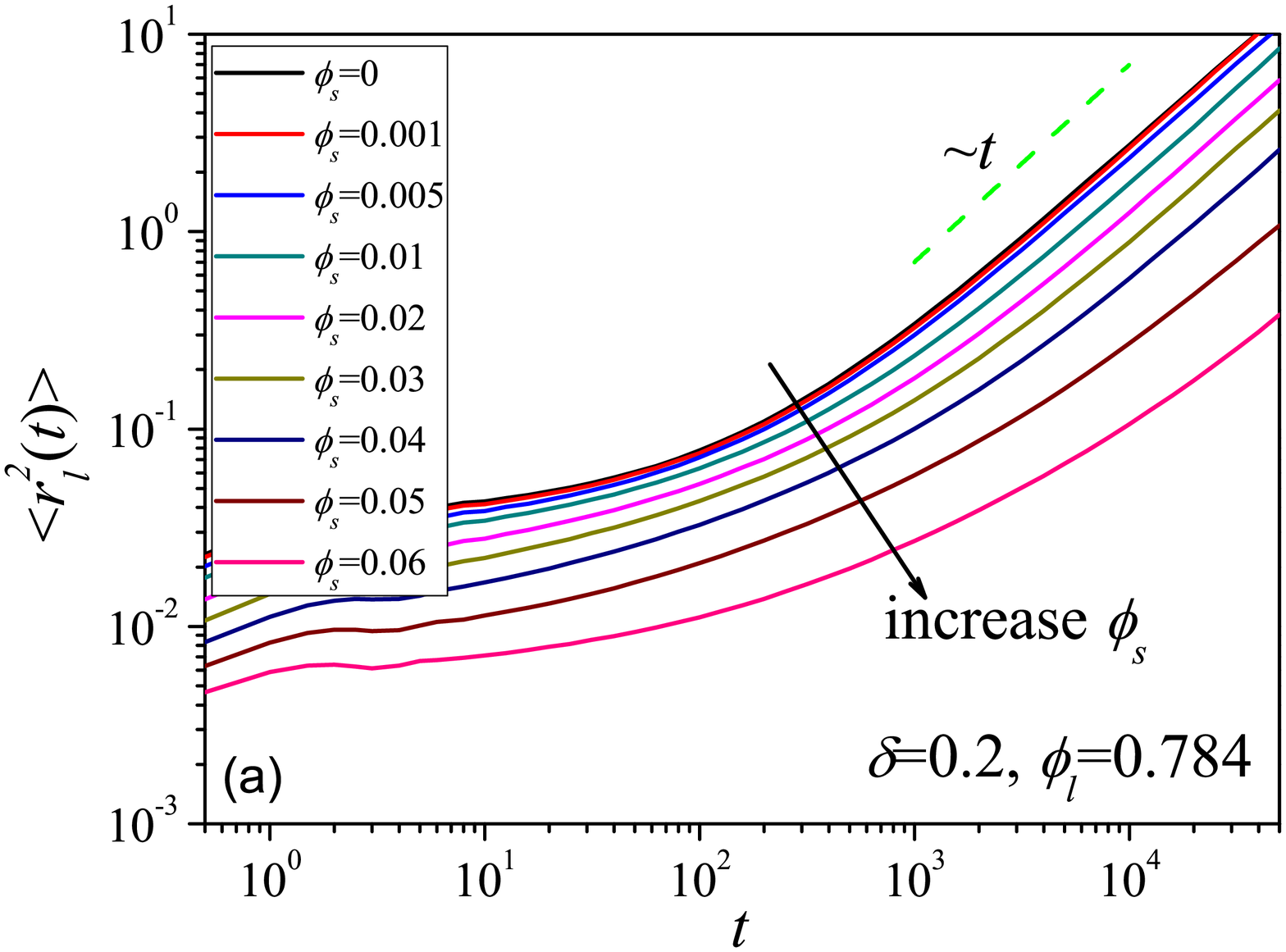}
 \includegraphics[angle=0,width=0.5\textwidth]{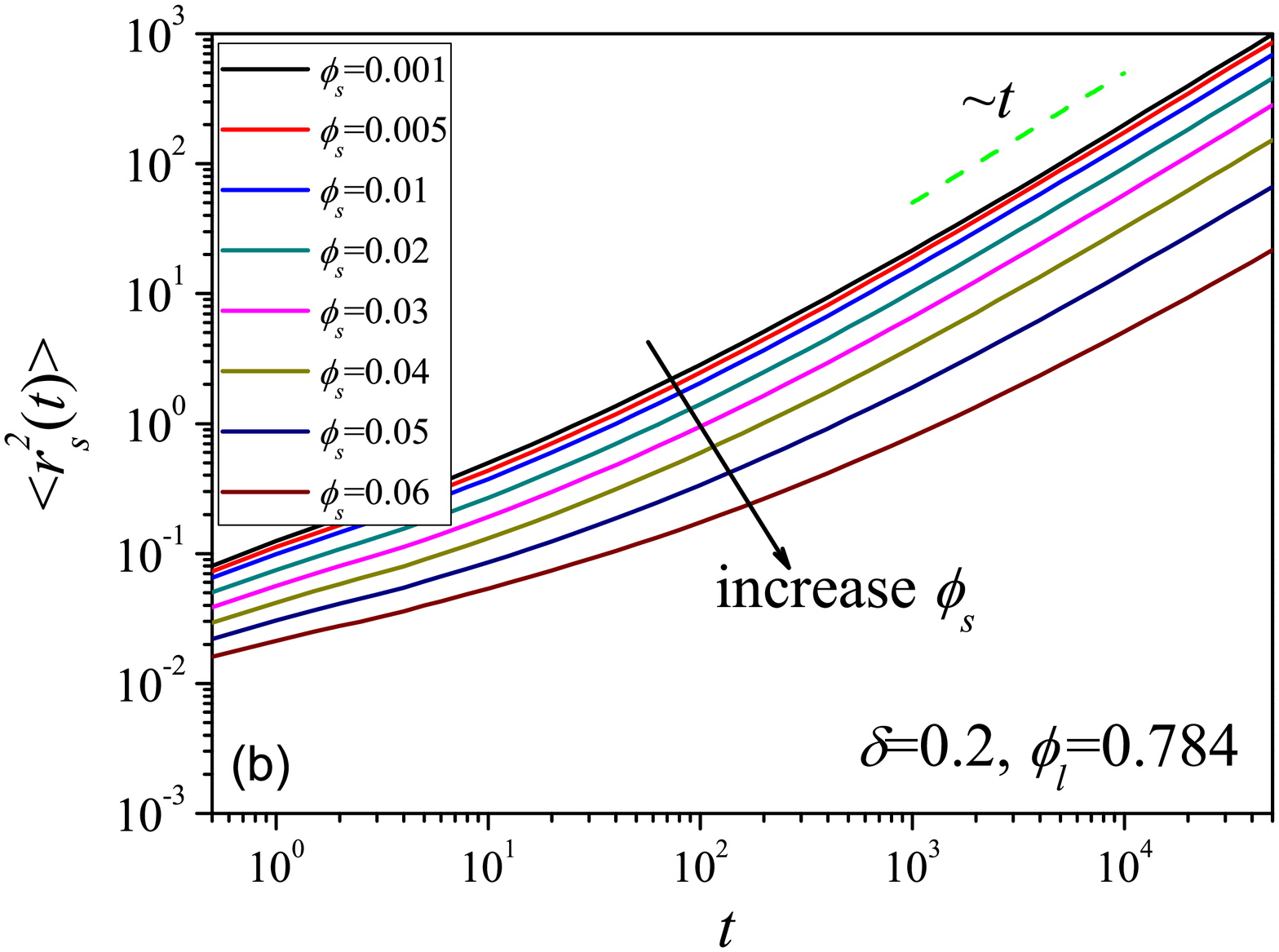}
 \caption{$\phi_{s}$ evolution of the mean squared displacements at $\phi_{l}=0.784$ and $\delta=0.2$ for (a) large and (b) small disks.}
\end{figure}

\begin{figure}[!htbp]
 \centering
 \includegraphics[angle=0,width=0.5\textwidth]{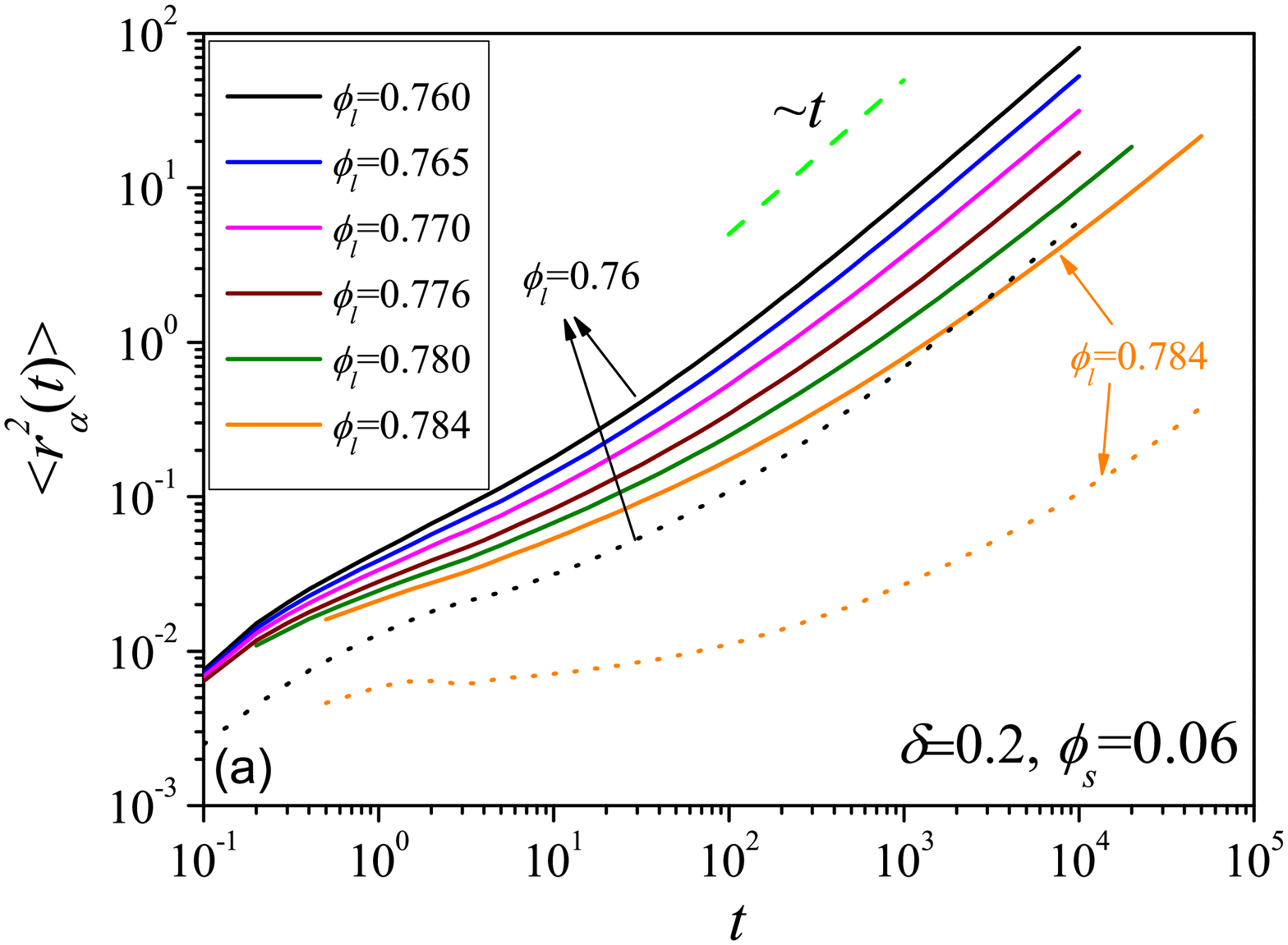}
 \includegraphics[angle=0,width=0.5\textwidth]{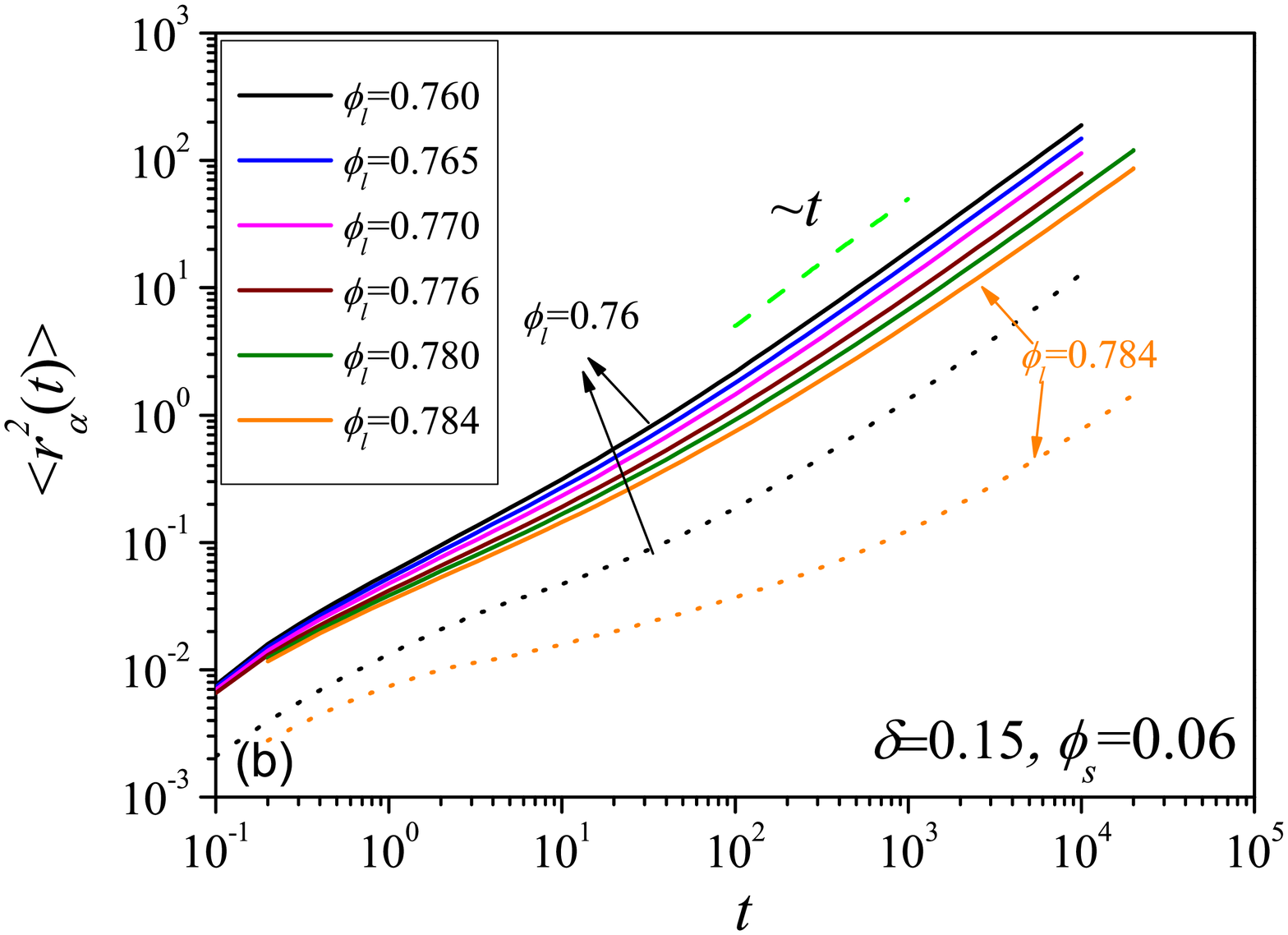}
 \caption{The mean squared displacements of large (dotted lines) and small (solid lines) disks at varying $\phi_{l}$ and $\phi_{s}=0.06$ for (a) $\delta=0.2$ and (b) $\delta=0.15$.}
\end{figure}

We now turn to the results concerning the dynamics of the binary hard-disk liquids as $\phi_{s}$ gradually increases. To this end, we have studied the self-intermediate scattering functions and the mean squared displacements (MSD) $<r^{2}_{\alpha}(t)>=\frac{1}{N_{\alpha}}<\sum_{j=1}^{N_{\alpha}}|\textbf{r}_{j}(t)-\textbf{r}_{j}(0)|^{2}>$, focusing on their $\phi_{s}$ evolution. We note that Moreno and Colmenero~\cite{Moreno1, Moreno2} have studied the temperature dependence of the mean squared displacements and the density-density correlators in soft-sphere mixtures with several specific compositions. They have found several striking dynamic features (as described in the introduction part) at varying temperature or wave vector due to different mechanisms of dynamic arrest for both species. Here, we focus on the change of the dynamics as $\phi_{s}$ increases. As mentioned above, the increase of $\phi_{s}$ leads to the enhancement of depletion effects for the large disks and bulk-like effects for the small disks. This can be more clearly revealed in Fig. 7, where particle trajectories for both species are presented during a time interval of $\tau_{l}$ (defined as $F_{s}^{l}(q_{p},t=\tau_l)=1/e$) for three $\phi_{l}$ values. Obviously, the small disks can only move within the voids left by the large disks (comparing different species' trajectories in Fig. 7), i.e., they are under confinement. It is also seen in Fig. 7 that the bulk-like effects for the small disks enhance as $\phi_{s}$ increases due to the increasing neighboring small particles. In addition, it seems that relaxation processes for both species are not uniform in space and that the region where the large disks move more slowly corresponds to the place where there are stronger confinement effects for the small disks. Therefore, we expect that some dynamic features observed in Ref.~\cite{Moreno1, Moreno2, Voigtmann3} will also emerge in our study as $\phi_{s}$ increases.

We first consider effects of adding small disks on the relaxation process of the large disks. In Fig. 8, the change of the self-intermediate scattering functions with $\phi_{s}$ is presented for the large disks. We observe two different relaxation scenarios for small (or moderate) and large $\phi_{l}$ values. For small $\phi_{l}$ values (e.g., $\phi_{l}=0.77$ in Fig. 8(a)), the time scales of $F_{s}^{l}(q_{p},t)$ for both initial and final decay increase monotonically as $\phi_{s}$ increases. This means that the increase of the total area fractions at small $\phi_{l}$ leads to a gradually slowing down of the dynamics for the large disks. When entering the regime of larger $\phi_{l}$ values, by contrast, another different relaxation scenario occurs, as shown in Fig. 8(b). Specifically, when $\phi_{s}$ increases from $0$ to $0.06$, the initial part of the structural relaxation for the large disks slows down and the plateau value at intermediate times increases, but the time scale for the final decay first decreases and then increases, exhibiting ``reentrant'' behavior. As one start tracing the glass-transition line at the highest density, the non-monotonic $\phi_{s}$ dependence of $\tau_{l}$ at $\phi_{l}=0.784$ means that the glass transition point for the large disks first becomes large and then shifts to smaller values upon increasing $\phi_{s}$, which is reminiscent of the ``reentrant'' glass transition observed in colloid-polymer mixtures (short-ranged attractive colloids)~\cite{Poon, Sciortino3} and the``inverted'' glass-transition curve proposed recently in a MCT study of highly size-asymmetric binary hard spheres~\cite{Voigtmann4}. Thus, a crossover for the structural relaxation of the large disks exists when the area fraction of the large disks is increased from small to large values. Turning to the relaxation process of the small disks themselves, several interesting features are also found as $\phi_{s}$ varies. It is seen in Fig. 9 that the shape of $F_{s}^{s}(q_{p},t)$ is concave at small $\phi_{s}$ values and becomes more convex as $\phi_{s}$ increases and that logarithmic decay also occurs for specific $\phi_{s}$ values. These dynamic features are strongly reminiscent of the findings in Ref.~\cite{Moreno1, Moreno2}. However, differing from the large disks, no crossover behavior occurs in the relaxation process of the small disks. To better characterize the $\phi_{s}$ evolution of the relaxation process at different $\phi_{l}$ values for both species, we plot relaxation times for both species (defined as $F_{s}^{\alpha}(q_{p},t=\tau_{\alpha})=1/e$) as a function of $\phi_{s}$ in Fig. 10. Clearly, $\tau_{s}$ increases monotonically at all $\phi_{l}$ values as $\phi_{s}$ increases. For the large disks, the monotonic change of $\tau_{l}$ with $\phi_{s}$ at small $\phi_{l}$ values crosses over into the non-monotonic behavior at large $\phi_{l}$ values. Moreover, we note that enhancing the size disparity (as $\delta$ decreases) leads to a drop for both $\tau_{l}$ and $\tau_{s}$ at fixed $\phi_{l}$ and $\phi_{s}$ values and does not alter the monotonic change of $\tau_{s}$ and the crossover of $\tau_{l}$ as $\phi_{s}$ increases, as shown in the insets of Fig. 10.

Thus, we have demonstrated that adding small disks can affect the relaxation process of both species in quantitative and qualitative ways. $\phi_{s}$ evolution of the relaxation process for the large disks is different for small and large $\phi_{l}$ values and several striking dynamic features for the small disks have been identified as $\phi_{s}$ varies. The influence of the small disks on the mean squared displacements can also be strong. In the first, we find from Fig. 11 that diffusion for both species slows down as $\phi_{s}$ increases, which is consistent with experimental results of Ref.~\cite{Imhof2} and implies that diffusion of both species is suppressed with increasing $\phi_{s}$. Moreover, we observe in Fig. 12 that long-time diffusion of the small disks displays power-law-like behavior (e.g., $<r^{2}_{s}(t)>\varpropto t^{\mu}$ with $\mu<1$ at $\phi_{l}=0.784$ for $t=10^{3}-10^{4}$) at sufficiently large $\phi_{s}$ values, in good agreement with Ref.~\cite{Voigtmann3}, and that the exponent $\mu$ decreases with increasing $\phi_{l}$. This anomalous, power-law-like diffusion of small disks implies that a glass transition for the large disks and a localization transition for the small disks can occur at sufficiently large $\phi_{s}$ values in our system. Furthermore, the power-law-like behavior becomes less evident for a smaller $\delta$ value (comparing Figs. 12(a) and 12(b)), indicating that the double-transition scenario occurs at larger $\phi_{s}$ values as the size disparity enhances.

\section{Conclusions}

In summary, we have investigated the structure, the compressibility factor and the dynamics of highly size-asymmetric binary hard-disk liquids. We have demonstrated that the addition of the small disks will not perturb the structure of the glass-forming liquids at the static pair level, but the higher-order static correlations can be strongly influenced, suggesting importance of higher-order correlations in understanding dynamic arrest of highly size-asymmetric binary mixtures. The compressibility factor of the system changes non-monotonically upon increasing the area fraction of the small disks and separating different contributions can rationalize this phenomenon. Adding a much smaller component can influence dynamics of the system in quantitative and qualitative ways. For the large disks, the structural relaxation time exhibits monotonic change with the area fraction of the small disks at low and moderate area fractions of the large disks. ``Reentrant'' behavior for the relaxation of the large disks is displayed at sufficiently high area fractions of the large disks, which strongly resembles the reentrant glass transition in short-ranged attractive colloids and the ``inverted'' glass transition in binary hard spheres with large size disparity. By tuning the area fraction of the small disks, relaxation process for the small disks shows concave-to-convex crossover and logarithmic decay behavior, as found in other binary mixtures. Diffusion of both species is suppressed by adding small disks, and in particular, the long-time diffusion for the small disks shows power-law-like behavior at sufficiently large $\phi_{s}$ values, which implies precursors of a glass transition for the large disks and a localization transition for the small disks. Our results can help to better understand dynamic arrest in highly size-asymmetric binary mixtures.

\begin{acknowledgments}
We thank Professors D.~Frenkel, M.~Miller, M.~Dijkstra, Z.~G.~Wang and Th.~Voigtmann for helpful discussions. This work is subsidized by the National Basic Research Program of China (973 Program, 2012CB821500), and supported by the National Natural Science Foundation of China (21074137, 50930001) programs and the fund for Creative Research Groups (50921062).
\end{acknowledgments}


\end{document}